# Title: The First Pulse of the Extremely Bright GRB 130427A: A Test Lab for Synchrotron Shocks

**Authors:** R. Preece[1*], J. Michael Burgess[2*], A. von Kienlin[3*], P. N. Bhat[2], M. S. Briggs[2], D. Byrne[4], V. Chaplin[2], W. Cleveland[5], A. C. Collazzi[6,7], V. Connaughton[2], A. Diekmann[8], G. Fitzpatrick[4], S. Foley[4,3], M. Gibby[8], M. Giles[8], A. Goldstein[6,7], J. Greiner[3], D. Gruber[3], P. Jenke[2], R. M. Kippen[9], C. Kouveliotou[6], S. McBreen[4,3], C. Meegan[2], W. S. Paciesas[5], V. Pelassa[2], D. Tierney[4], A. J. van der Horst[10], C. Wilson-Hodge[6], S. Xiong[2], G. Younes[5,6], H.-F. Yu[3], M. Ackermann[11], M. Ajello[12], M. Axelsson[13,14,15], L. Baldini[16], G. Barbiellini[17,18], M. G. Baring[19], D. Bastieri[20,21], R. Bellazzini[22], E. Bissaldi[23], E. Bonamente[24,25], J. Bregeon[22], M. Brigida[26,27], P. Bruel[28], R. Buehler[11], S. Buson[20,21], G. A. Caliandro[29], R. A. Cameron[30], P. A. Caraveo[31], C. Cecchi[24,25], E. Charles[30], A. Chekhtman[32], J. Chiang[30], G. Chiaro[21], S. Ciprini[33,34], R. Claus[30], J. Cohen-Tanugi[35], L. R. Cominsky[36], J. Conrad[37,14,38,39], F. D'Ammando[40], A. de Angelis[41], F. de Palma[26,27], C. D. Dermer[42*], R. Desiante[17], S. W. Digel[30], L. Di Venere[30], P. S. Drell[30], A. Drlica-Wagner[30], C. Favuzzi[26,27], A. Franckowiak[30], Y. Fukazawa[43], P. Fusco[26,27], F. Gargano[27], N. Gehrels[44], S. Germani[24,25], N. Giglietto[26,27], F. Giordano[26,27], M. Giroletti[40], G. Godfrey[30], J. Granot[45], I. A. Grenier[46], S. Guiriec[44,7], D. Hadasch[29], Y. Hanabata[43], A. K. Harding[44], M. Hayashida[30,47], S. Iyyani[14,15,37], T. Jogler[30], G. Jóannesson[48], T. Kawano[43], J. Knödlseder[49,50], D. Kocevski[30], M. Kuss[22], J. Lande[30], J. Larsson[15,14], S. Larsson[37,14,13], L. Latronico[51], F. Longo[17,18], F. Loparco[26,27], M. N. Lovellette[42], P. Lubrano[24,25], M. Mayer[11], M. N. Mazziotta[27], P. F. Michelson[30], T. Mizuno[52], M. E. Monzani[30], E. Moretti[15,14], A. Morselli[53], S. Murgia[30], R. Nemmen[44], E. Nuss[35], T. Nymark[15,14], M. Ohno[54], T. Ohsugi[52], A. Okumura[30,55], N. Omodei[30*], M. Orienti[40], D. Paneque[56,30], J. S. Perkins[44,57,58], M. Pesce-Rollins[22], F. Piron[35], G. Pivato[21], T. A. Porter[30], J. L. Racusin[44], S. Rainò[26,27], R. Rando[20,21], M. Razzano[22,59], S. Razzaque[60], A. Reimer[23,30], O. Reimer[23,30], S. Ritz[59], M. Roth[61], F. Ryde[15], A. Sartori[31], J. D. Scargle[62], A. Schulz[11], C. Sgrò[22], E. J. Siskind[63], G. Spandre[22], P. Spinelli[26,27], D. J. Suson[64], H. Tajima[30,55], H. Takahashi[43], J. G. Thayer[30], J. B. Thayer[30], L. Tibaldo[30], M. Tinivella[22], D. F. Torres[29,65], G. Tosti[24,25], E. Troja[44,66], T. L. Usher[30], J. Vandenbroucke[30], V. Vasileiou[35], G. Vianello[30,67], V. Vitale[53,68], M. Werner[23], B. L. Winer[69], K. S. Wood[42], S. Zhu[66].

**Affiliations:**

[1]Department of Space Science, University of Alabama in Huntsville, Huntsville, AL 35899, USA
[2]Center for Space Plasma and Aeronomic Research (CSPAR), University of Alabama in Huntsville, Huntsville, AL 35899, USA
[3]Max-Planck Institut für extraterrestrische Physik, 85748 Garching, Germany
[4]University College Dublin, Belfield, Dublin 4, Ireland
[5]Universities Space Research Association (USRA), Columbia, MD 21044, USA
[6]NASA Marshall Space Flight Center, Huntsville, AL 35812, USA
[7]NASA Postdoctoral Program Fellow, USA
[8]Jacobs Technology, Huntsville, AL 35806, USA
[9]Los Alamos National Laboratory, Los Alamos, NM 87545, USA
[10]Astronomical Institute "Anton Pannekoek" University of Amsterdam, Postbus 94249 1090 GE Amsterdam, Netherlands
[11]Deutsches Elektronen Synchrotron DESY, D-15738 Zeuthen, Germany

[12]Space Sciences Laboratory, 7 Gauss Way, University of California, Berkeley, CA 94720-7450, USA
[13]Department of Astronomy, Stockholm University, SE-106 91 Stockholm, Sweden
[14]The Oskar Klein Centre for Cosmoparticle Physics, AlbaNova, SE-106 91 Stockholm, Sweden
[15]Department of Physics, Royal Institute of Technology (KTH), AlbaNova, SE-106 91 Stockholm, Sweden
[16]Università di Pisa and Istituto Nazionale di Fisica Nucleare, Sezione di Pisa I-56127 Pisa, Italy
[17]Istituto Nazionale di Fisica Nucleare, Sezione di Trieste, I-34127 Trieste, Italy
[18]Dipartimento di Fisica, Università di Trieste, I-34127 Trieste, Italy
[19]Rice University, Department of Physics and Astronomy, MS-108, P. O. Box 1892, Houston, TX 77251, USA
[20]Istituto Nazionale di Fisica Nucleare, Sezione di Padova, I-35131 Padova, Italy
[21]Dipartimento di Fisica e Astronomia "G. Galilei'", Università di Padova, I-35131 Padova, Italy
[22]Istituto Nazionale di Fisica Nucleare, Sezione di Pisa, I-56127 Pisa, Italy
[23]Institut für Astro- und Teilchenphysik and Institut für Theoretische Physik, Leopold-Franzens-Universität Innsbruck, A-6020 Innsbruck, Austria
[24]Istituto Nazionale di Fisica Nucleare, Sezione di Perugia, I-06123 Perugia, Italy
[25]Dipartimento di Fisica, Università degli Studi di Perugia, I-06123 Perugia, Italy
[26]Dipartimento di Fisica "M. Merlin" dell'Università e del Politecnico di Bari, I-70126 Bari, Italy
[27]Istituto Nazionale di Fisica Nucleare, Sezione di Bari, 70126 Bari, Italy
[28]Laboratoire Leprince-Ringuet, Ècole polytechnique, CNRS/IN2P3, Palaiseau, France
[29]Institut de CiËncies de l'Espai (IEEE-CSIC), Campus UAB, 08193 Barcelona, Spain
[30]W. W. Hansen Experimental Physics Laboratory, Kavli Institute for Particle Astrophysics and Cosmology, Department of Physics and SLAC National Accelerator Laboratory, Stanford University, Stanford, CA 94305, USA
[31]INAF-Istituto di Astrofisica Spaziale e Fisica Cosmica, I-20133 Milano, Italy
[32]Center for Earth Observing and Space Research, College of Science, George Mason University, Fairfax, VA 22030, resident at Naval Research Laboratory, Washington, DC 20375, USA
[33]Agenzia Spaziale Italiana (ASI) Science Data Center, I-00044 Frascati (Roma), Italy
[34]Istituto Nazionale di Astrofisica - Osservatorio Astronomico di Roma, I-00040 Monte Porzio Catone (Roma), Italy
[35]Laboratoire Univers et Particules de Montpellier, Universitè Montpellier 2, CNRS/IN2P3, Montpellier, France
[36]Department of Physics and Astronomy, Sonoma State University, Rohnert Park, CA 94928-3609, USA
[37]Department of Physics, Stockholm University, AlbaNova, SE-106 91 Stockholm, Sweden
[38]Royal Swedish Academy of Sciences Research Fellow, funded by a grant from the K. A. Wallenberg Foundation
[39]The Royal Swedish Academy of Sciences, Box 50005, SE-104 05 Stockholm, Sweden
[40]INAF Istituto di Radioastronomia, 40129 Bologna, Italy
[41]Dipartimento di Fisica, Università di Udine and Istituto Nazionale di Fisica Nucleare, Sezione di Trieste, Gruppo Collegato di Udine, I-33100 Udine, Italy
[42]Space Science Division, Naval Research Laboratory, Washington, DC 20375-5352, USA
[43]Department of Physical Sciences, Hiroshima University, Higashi-Hiroshima, Hiroshima 739-

8526, Japan
[44]NASA Goddard Space Flight Center, Greenbelt, MD 20771, USA
[45]Department of Natural Sciences, The Open University of Israel, 1 University Road, POB 808, Ra'anana 43537, Israel
[46]Laboratoire AIM, CEA-IRFU/CNRS/Universitè Paris Diderot, Service d'Astrophysique, CEA Saclay, 91191 Gif sur Yvette, France
[47]Department of Astronomy, Graduate School of Science, Kyoto University, Sakyo-ku, Kyoto 606-8502, Japan
[48]Science Institute, University of Iceland, IS-107 Reykjavik, Iceland
[49]CNRS, IRAP, F-31028 Toulouse cedex 4, France
[50]GAHEC, Universitè de Toulouse, UPS-OMP, IRAP, Toulouse, France
[51]Istituto Nazionale di Fisica Nucleare, Sezione di Torino, I-10125 Torino, Italy
[52]Hiroshima Astrophysical Science Center, Hiroshima University, Higashi-Hiroshima, Hiroshima 739-8526, Japan
[53]Istituto Nazionale di Fisica Nucleare, Sezione di Roma “Tor Vergata”, I-00133 Roma, Italy
[54]Institute of Space and Astronautical Science, JAXA, 3-1-1 Yoshinodai, Chuo-ku, Sagamihara, Kanagawa 252-5210, Japan
[55]Solar-Terrestrial Environment Laboratory, Nagoya University, Nagoya 464-8601, Japan
[56]Max-Planck-Institut für Physik, D-80805 München, Germany
[57]Department of Physics and Center for Space Sciences and Technology, University of Maryland Baltimore County, Baltimore, MD 21250, USA
[58]Center for Research and Exploration in Space Science and Technology (CRESST) and NASA Goddard Space Flight Center, Greenbelt, MD 20771, USA
[59]Santa Cruz Institute for Particle Physics, Department of Physics and Department of Astronomy and Astrophysics, University of California at Santa Cruz, Santa Cruz, CA 95064, USA
[60]University of Johannesburg, Department of Physics, University of Johannesburg, Auckland Park 2006, South Africa,
[61]Department of Physics, University of Washington, Seattle, WA 98195-1560, USA
[62]Space Sciences Division, NASA Ames Research Center, Moffett Field, CA 94035-1000, USA
[63]NYCB Real-Time Computing Inc., Lattingtown, NY 11560-1025, USA
[64]Department of Chemistry and Physics, Purdue University Calumet, Hammond, IN 46323-2094, USA
[65]Instituciò Catalana de Recerca i Estudis Avançats (ICREA), Barcelona, Spain
[66]Department of Physics and Department of Astronomy, University of Maryland, College Park, MD 20742, USA
[67]Consorzio Interuniversitario per la Fisica Spaziale (CIFS), I-10133 Torino, Italy
[68]Dipartimento di Fisica, Università di Roma “Tor Vergata”, I-00133 Roma, Italy
[69]Department of Physics, Center for Cosmology and Astro-Particle Physics, The Ohio State University, Columbus, OH 43210, USA

*Correspondence to: R. Preece, preecer@uah.edu; J. Michael Burgess, James.Burgess@uah.edu; C. D. Dermer, charles.dermer@nrl.navy.mil; N. Omodei, nicola.omodei@stanford.edu; A. von Kienlin, azk@mpe.mpg.de.

**Abstract**: Gamma-ray burst (GRB) 130427A is one of the most energetic GRBs ever observed. The initial pulse up to 2.5 s is possibly the brightest well-isolated pulse observed to date. A fine time resolution spectral analysis shows power-law decays of the peak energy from the onset of the pulse, consistent with models of internal synchrotron shock pulses. However, a strongly correlated power-law behavior is observed between the luminosity and the spectral peak energy that is inconsistent with curvature effects arising in the relativistic outflow. It is difficult for any of the existing models to account for all of the observed spectral and temporal behaviors simultaneously.

**Main Text:** In the context of gamma-ray bursts, GRB 130427A, which triggered the Gamma-ray Burst Monitor (GBM) (*1*) on the Fermi Gamma-ray Space Telescope on 2013-Apr-27 at $T_0$ = 07:47:06.42 UTC (*2 - 4*) is an extreme case. The peak flux on the 64 ms timescale is 1300 ± 100 photons $s^{-1}$ $cm^{-2}$ in the 10 – 1000 keV range and the fluence, integrated over the same energy range and a total duration of approximately 350 s, is (2.4 ± 0.1) × $10^{-3}$ erg $cm^{-2}$. The longest continuously running GRB detector, Konus on the Wind spacecraft, has been observing the entire sky for nearly 18 years and only one burst had a larger peak flux, by ~30% (GRB 110918A) (*5*). GRB 130427A is the most fluent burst in the era starting with the 1991 launch of the Burst And Transient Source Experiment (BATSE) on the Compton Gamma-Ray Observatory. Finally, the energy of the spectral peak in the first time bin ($T_0$ –0.1 to 0.0 s), 5400 ± 1500 keV, is the second highest ever recorded (*6*).

The initial pulse (Fig. 1), lasting up to 2.5 s after the trigger, stands on its own as being so bright (170 ± 10 ph $s^{-1}$ $cm^{-2}$ peak flux for 10 – 1000 keV in the 64 ms time bin at $T_0$ +0.51 s) as to be ranked among the 10 brightest GBM or BATSE bursts (*7 - 9*). The brightness allows us to track the spectral evolution of the rising portion of a well-separated pulse with unprecedented detail (*10*). Evident in the GBM low-energy light curve (Fig. 1; as well as the 15 – 350 keV light curve presented in (*11*)) are fluctuations starting at around 1 s that are not present at higher energies. If these represent additional low-energy pulses, their presence clearly does not dominate the analyses presented below.

Past studies of time-resolved spectra of simple pulses in GRBs indicate that there are broadly two classes of spectral evolution. These are called 'hard-to-soft' and 'tracking' pulses (*12, 13*), depending on whether the energy of the peak in the $\nu F_\nu$ spectrum (generically called $E_{\text{peak}}$ herein) monotonically decays independently of the flux evolution or else generally follows the rise and fall of the flux. Typically, there are at most one or two spectra available for fitting during the rising portion of the flux history. What makes this event unique is that there are roughly 6 time bins with excellent counts statistics before the peak in the 10 – 1000 keV flux.

As seen in Fig. 1, there is a clear trend in the individual detector's light curves: the > 20 MeV Fermi Large Area Telescope (LAT) low-energy (LLE) (*14, 15*) light curve peaks before the GBM trigger time ($T_0$), while the GBM bismuth germanate (BGO) detector #1 (300 keV – 45 MeV) and sodium iodide (NaI) detector #6 (8 – 300 keV) peak at successively later times. To quantify this, we performed an energy-dependent pulse lag analysis using a Discrete Cross Correlation Function (DCCF) and obtain the time lags $\tau$ (*16*) between the highest energy LAT LLE light curve and light curves at several selected energy ranges in the GBM NaI and BGO detectors (Fig. 1 – *inset*). We find good agreement between the expected lag behavior and the pulse width model $W(E) \propto E^{\alpha}$ (*17*), obtaining a fitted value for $\alpha$ = –0.27 ± 0.03 (*18*). This model was previously fit to 400 pulses from 41 BATSE GRBs (*17*); an average value of $\alpha$ = –0.41 was found. Synchrotron shock model simulations made by (*19*) found $\alpha$ ~ –0.4 for pulses of 2 – 10 s duration but $\alpha$ > ~ –0.2 for pulses of 0.1 – 1 s. Three LAT photons with energies greater than 100 MeV are clustered in coincidence with the LLE peak, and so may arise by the same mechanism.

Although most GRB spectra are well fit by the smoothly joined broken power-law function of Band et al. (*20*), in some cases the simultaneous fit of a Band function together with an additional blackbody component is significantly better statistically (*6, 21, 22*). Burgess et al. (*23, 24*) also noted the requirement of an additional blackbody component, but replaced the phenomenological Band function with a physically-motivated synchrotron function. We present two separate time-resolved spectral analyses for the first 2.5 s of GRB 130427A, with comparable goodness of fit: the Burgess et al. synchrotron function plus blackbody, and the Band function (*18*). A blackbody component is not required when the more flexible Band function alone is used. Although the time evolution of $E_{\mathrm{peak}}$ as determined by Band function fits is consistent with a single power-law (with an index of $-0.96 \pm 0.02$), the evolution of the synchrotron peak energy is not (Fig. 2). A broken power-law fit is better constrained and shows a shallower decay before the pulse peak, with an index of $-0.4 \pm 0.2$ during the rising phase and $-1.17 \pm 0.05$ during the decaying phase and a fitted break time at $T_0 + 0.28 \pm 0.08$ s, or ~ 0.2 s before the pulse peak in the 10 – 1000 keV flux. Both fitted indices during the decay phase are consistent with the –1 power law index expected from standard fireball curvature effects (*25, 26*). The shallower spectral peak decay index prior to the light curve decay phase has a natural explanation in the context of the pulse being driven by a shock between thick colliding shells (*19*).

Two-component models including a thermal contribution (*27*) constrain the value of the photospheric radius using the blackbody flux and temperature ($kT$ - see Table S1). Comparing with the flux of the dominant non-thermal spectral component then permits determination of the Lorentz factor at the photosphere ($\Gamma_{\mathrm{ph}}$) (*28*). As shown in Fig. 3, the minimum value of the photospheric bulk Lorentz factor $\Gamma_{\mathrm{ph}}$ starts out at 500 and monotonically decreases to ~100 over the duration of the pulse (similar to behavior observed in GRB 110721A) (*29*). Internal shocks require higher Lorentz factors at later times. However, this might still be consistent with the monotonically decreasing $\Gamma_{\mathrm{ph}}$ if the outflowing shell that produces this photospheric component produced the non-thermal triggering pulse by colliding with a slower and slightly earlier ejected shell that did not produce detectable photospheric emission. Otherwise, the observed behavior would favor magnetic reconnection models or mini-jets (*30, 31*), which abandon a simple spherical geometry.

Using the measured redshift of $z = 0.34$ (*32*), the host rest-frame luminosity and synchrotron peak energies are calculated and the decay phase apparent isotropic luminosity $L$-$E_{\mathrm{peak}}$ correlation is fit with a power-law index of $1.43 \pm 0.04$ (Fig. 4). A theoretical analysis of high-latitude curvature radiation produced in relativistic shell collisions of spherical blast waves shows that $L \propto E_{\mathrm{peak}}^3$ during the decay phase of a pulse (*25, 26*), contrary to the behavior shown in Fig. 4. In a picture of an expanding fluid element rather than a colliding shell, synchrotron emission by electrons with characteristic energy $\gamma_e$ obeys the relations $E_{\mathrm{peak}} \propto \Gamma B \gamma_e^2$ and $L \propto \Gamma^2 B^2 \gamma_e^2$, for $\gamma_e$ and $B$ both in the jet frame. In the optically-thin coasting phase of the outflow, the bulk Lorentz factor $\Gamma$ is constant. Naively assuming that the magnetic flux is frozen in the flow ($BR^2 \propto$ *const.* in the co-moving frame, where $R$ is the comoving emission region radius), then adiabatic losses of the electrons imply $\gamma_e \propto R^{-1}$. A short calculation then gives $L \propto E_{\mathrm{peak}}^{3/2}$, which is consistent with the 1.43 index derived from the data. A constant expansion velocity

d$R$/d$t$ scenario predicts, however, $E_{\rm peak} \propto R^{-4} \propto t^{-4}$. Other jet-wind assumptions yield different correlations, e.g., for the deceleration epoch where d$\Gamma$/d$t$ < 0, or for radial field evolution appropriate for jet cores.

The isolated initial pulse of GRB 130427A is apparently unmodified by preceding engine activity or nascent external shock emission. Our analysis shows that there is good agreement between the pulse width as a function of energy and the expected lag, the characteristic energy has roughly a –1 power-law decay with time during the decaying phase, the temperature of the blackbody component implies a photospheric radius that is incompatible with the internal shock radius, and the apparent isotropic luminosity is related to the 3/2 power of the characteristic energy. It is a challenge to explain all these behaviors simultaneously.

**Acknowledgments:** The Fermi data are publically available at NASA's Fermi Science Support Center's website: http://fermi.gsfc.nasa.gov/ssc/.

The Fermi GBM collaboration acknowledges support for GBM development, operations and data analysis from NASA in the US and BMWi/DLR in Germany.

The Fermi LAT Collaboration acknowledges support from a number of agencies and institutes for both development and the operation of the LAT as well as scientific data analysis. These include NASA and DOE in the United States, CEA/Irfu and IN2P3/CNRS in France, ASI and INFN in Italy, MEXT, KEK, and JAXA in Japan, and the K. A. Wallenberg Foundation, the Swedish Research Council and the National Space Board in Sweden. Additional support from INAF in Italy and CNES in France for science analysis during the operations phase is also gratefully acknowledged.

**Supplementary Online Materials:**

SOM Text

References (*33* – *35*)

Table S1 (submitted as a separate file)

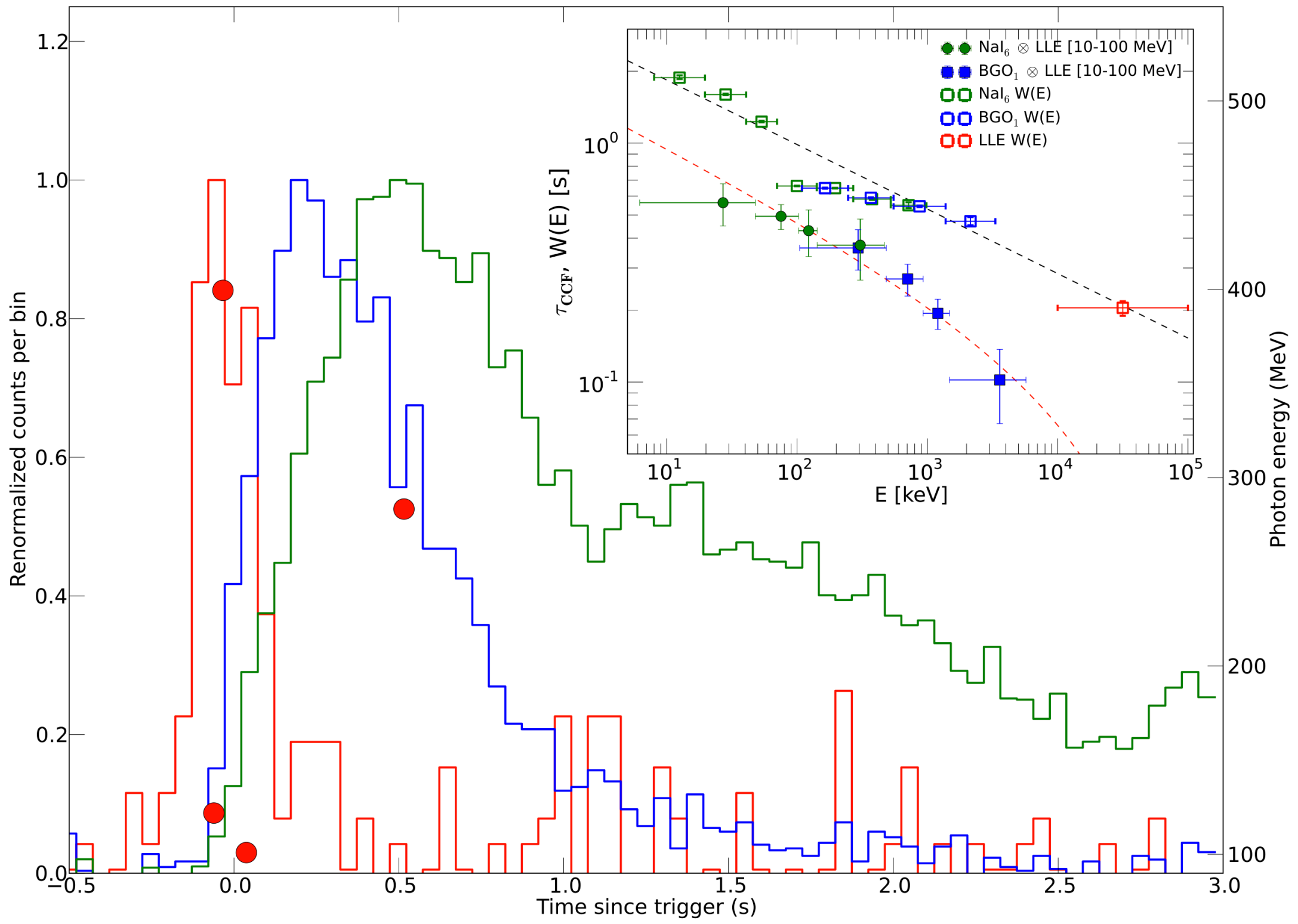

**Fig. 1.** The first 3 s of GRB 130427A. Shown are composite light curves for the three Fermi detector types (*green*: GBM NaI #6 [10 – 300 keV]; *blue*: GBM BGO #1 [300 keV – 45 MeV]; *red*: LAT LLE [> 20 MeV]). Each curve has been normalized so that their peak intensities match. High probability LAT photons > 100 MeV are indicated by circles (*right axis* – energy in MeV). (*Inset Figure*:) Lag analysis of the triggering pulse of GRB 130427A. Time lag $\tau$ (*filled symbols*) as determined by the DCCF analysis between the (10-100 MeV) LLE lightcurve and selected energy bands of the NaI (*green*) and BGO (*blue*) lightcurves. Also displayed are fitted pulse widths as a function of energy $W(E)$ (*hollow symbols*, in sec.) for several energy bands. The two dashed lines represent: 1), the best-fit power-law model ($\chi^2$ of 5.6 for 9 degrees of freedom) for $W(E)$ (*black*), and 2), the expected dependence of the time lag $\tau$ as a function of energy (*red*), assuming the same power-law index as in 1).

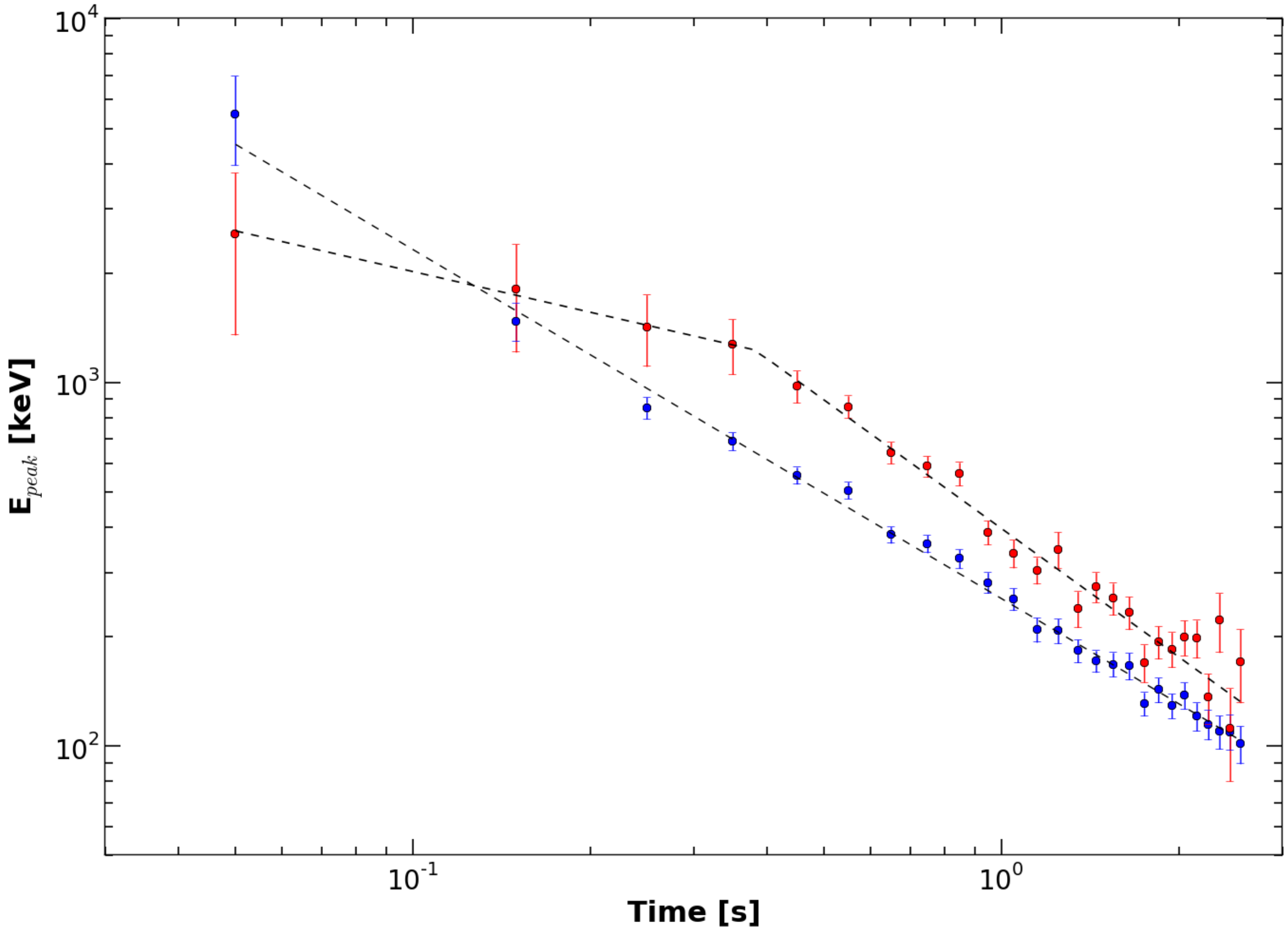

**Fig. 2.** The fitted Band function $E_{\mathrm{peak}}$ (*blue*) and synchrotron peak energies (*red*) as a function of time. The times are referenced from when the LLE light curve peaks 0.1 s before the trigger. A broken power-law fit to the red points is indicated by a dashed line (early time decay index of –0.4 ± 0.2, with a break at 0.38 ± 0.08 s, breaking to an index of –1.17 ± 0.05 with a $\chi^2 = 28$ for 22 degrees of freedom). We also show the Band function $E_{\mathrm{peak}}$ values for the same time intervals with a single fitted power law index (–0.96 ± 0.02 with a $\chi^2$ of 19 for 24 degrees of freedom).

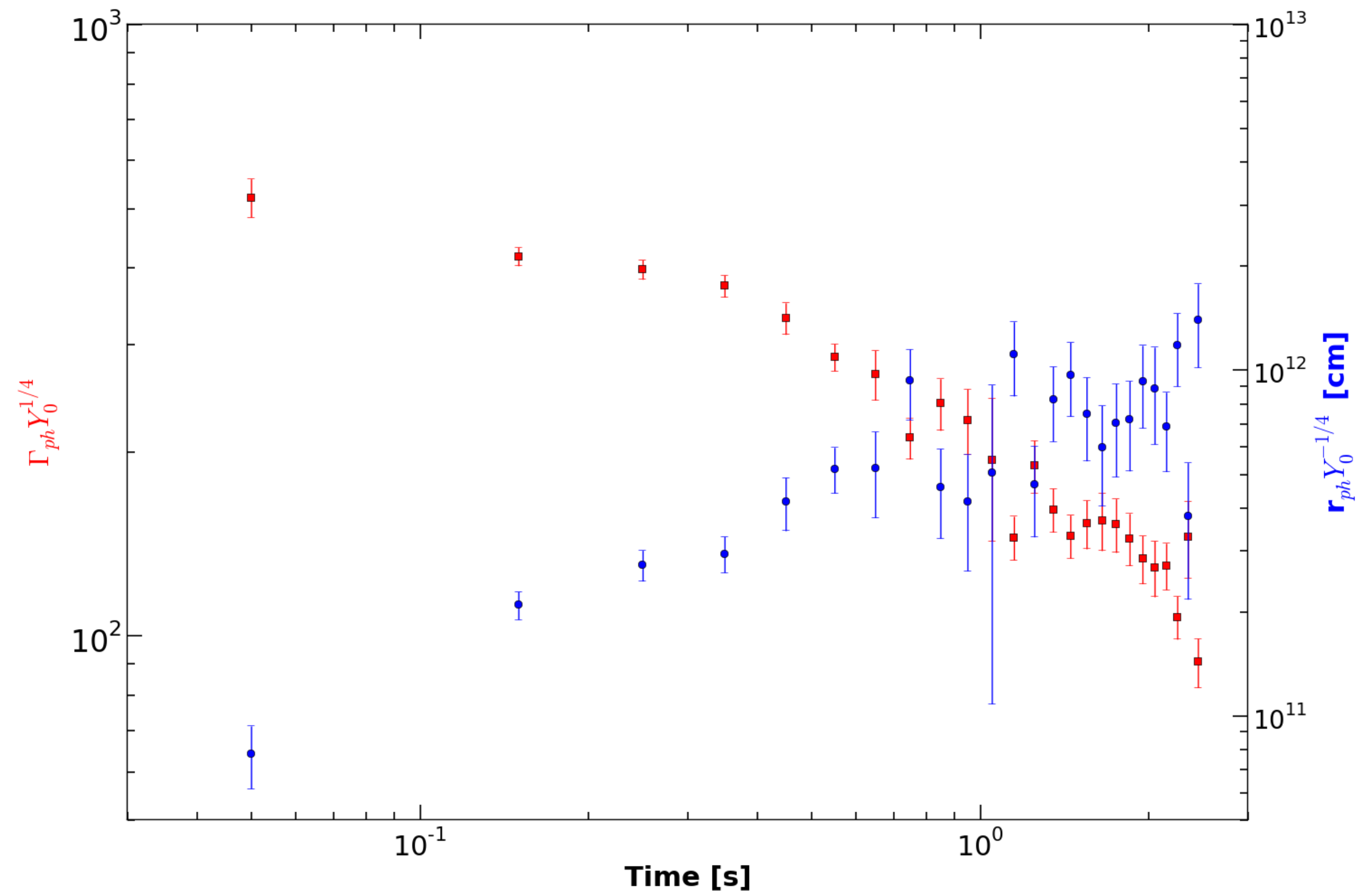

**Fig. 3.** Plot showing trends in the derived photospheric Lorentz factor (*red – left axis*) and radius (*blue – right axis*). The reference time is the same as in Fig. 2. We obtain both values from the instantaneous ratio of the observed blackbody component flux to the total flux, following Eq. 4 & 5 in (*28*) and assuming a value of $Y_0 = 1$ for the ratio between the total emitted thermal energy vs. the total energy emitted in gamma rays.

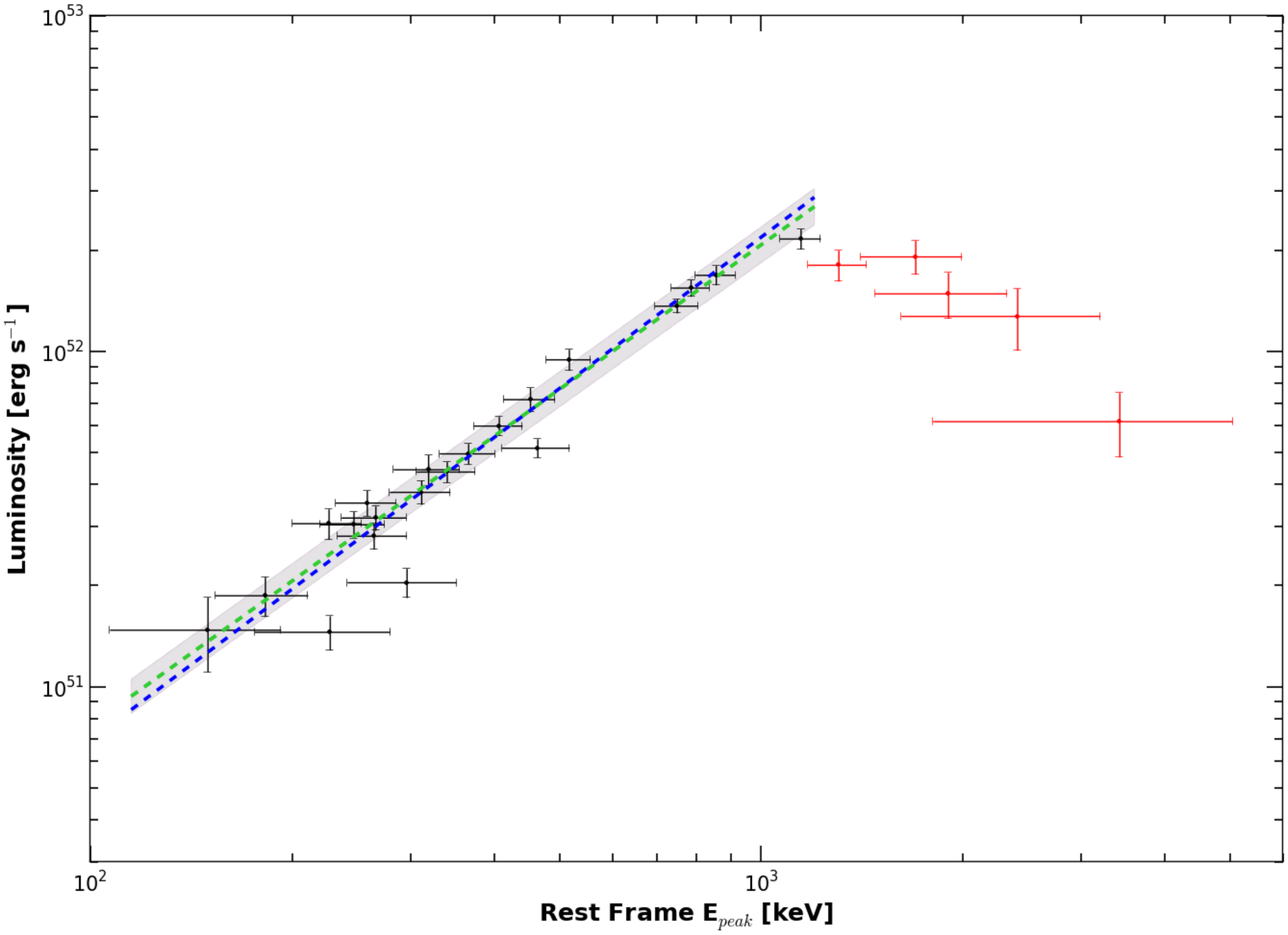

**Fig. 4.** Correlation between the GRB 130427A host rest frame synchrotron peak energy and isotropic luminosity during the rising phase (*red*) and decaying phase (*black*) of the triggering pulse. Time progresses approximately from right to left on the plot. The 1.43 ± 0.04 power-law index fit to the black points is shown in green (region of uncertainty in grey), while the 3/2 power-law from the magnetic flux-freezing calculation in the text is indicated in blue.

## Supplementary Online Materials:

### Spectral Analysis Method: Band Function Example

The GBM time-tagged event (TTE) data (*1*) and the LAT low-energy events (LLE) data (*14, 15*) were binned at 100 ms resolution, referenced from the GBM trigger, covering a continuous energy range from 8 keV to 100 MeV. We used GBM data for BGO #1, plus NaI #6, with source-to-zenith angles of 53.5° and 6.5° respectively, as determined for the best available CARMA radio source position (173.1367° RA, +27.6989° Dec ± 0.4", J2000) (*33*). Using the spectral analysis package RMFIT (*34*), we first fit the Band GRB function (*20*) to the entire interval from $T_0$ –0.1 to 2.5 s in order to determine the relative offset between the BGO and NaI rates, since there is typically an effective area correction of 10% between the BGO and any of the NaI detectors. We fix the effective area correction between the BGO and LLE rates to 1.0, as there is no known offset between the two and unfreezing the parameter drives it to large, unreasonable values. The best-fit Band parameter values are $E_{\rm peak}$ = 376 ± 7 keV, $\alpha_{\rm B}$ (the low-energy power law index) = –0.86 ± 0.01 and $\beta$ (the high-energy power law index) = –2.66 ± 0.035, with C-Stat = 599.65 for 377 degrees of freedom and poor residuals. With this spectrum as a seed, we then perform spectral fits for each of the time bins in the series. Examination of the results indicates that the $\beta$ parameter is undetermined in roughly half the spectra and when it could be determined by the fits, $\beta$ never deviated beyond one standard deviation from the value determined from the fit to the entire interval. We find $\beta$ to be consistent with a constant value of –2.66 and so the results shown are based on fixing $\beta$ to that value and fitting the other parameters. The spectral evolution of the $E_{\rm peak}$ parameter is evident in Fig. 2, falling smoothly from the onset of the pulse; the complete set of spectral fit parameters can be found in Table S1. Finally, the distribution in $\alpha_{\rm B}$ peaks very close to –0.66, or –2/3, the value expected for optically thin synchrotron emission.

If one allows for the evolution of a magnetic field in a fast-cooling synchrotron scenario, low-energy spectral indices of –0.8 are possible, as shown by (*35*). The fitted $\alpha_{\rm B}$ values during the brightest portion of the pulse fall in the –0.2 to –0.6 range, many sigma away from the desired –0.8 spectral index. It is usually the case that if one then imposes a low energy spectral index value (say, –0.8), a second component, such as a blackbody, is required to make up the difference. So, we would have to allow a blackbody component along with the quasi-Band spectrum, in this purely magnetic model. Interestingly, the Band $\alpha_{\rm B}$ does settle on an average that is nearly –0.8 during the later decay phase.

### Lag Analysis

An empirically-motivated function for the expected time lag between two pulses of energy $E_1$ and $E_2$ is given by $\tau(E_1, E_2) = t_{\rm rise}\,(E_1^{\alpha} - E_2^{\alpha})$, where their full widths at half maximum (FWHMs - in s) are related by the function $W(E) = W_0\,(E/E_0)^{\alpha}$, where $E_0$ = 1 keV (*17*). We fitted pulses at several energies with a lognormal function to obtain the FWHM in each energy band (Fig. 1 – *inset, hollow symbols*). These were then used to derive the parameters $W_0$ = 3.2 ± 0.5 s and $\alpha$ = –0.27 ± 0.03. We find good agreement between the above expected lag behavior, using our fitted value for $\alpha$ (Fig. 1 – *inset, dashed lines*) and the values of $t_{\rm rise}$ determined by the width fitting.

### Photospheric Radius and Lorentz Factor

Given the joint observations of the photospheric radius and Lorentz factor, how consistent are these with the standard internal shock model of GRBs? The radius $r_{\rm IS}$ of the corresponding internal shell collisions then should be $r_{IS} \sim c\,\Gamma^2\,\Delta t \sim 7.5\times10^{15}\,(\Gamma/500)^2$ cm >> $r_{ph} \sim 10^{12}$ cm, where $\Delta t \sim 1$ s is the pulse duration (see Fig. 1), which clearly is not the case, as the photospheric and synchrotron spectral components are so closely coupled in time. Such large radii are also required for the γ-ray emission region to be optically thin to γγ absorption, which would cut off the spectrum at lower energies than observed. Because the wind escaping the photosphere travels more slowly than light, the emission from the colliding shells would lag behind the photospheric radiation by $\sim\Delta t$. Finally, we note that $r_{ph}$ and $\Gamma$ scale with the ratio between the total gamma-ray energy and the thermal energy $Y_0$, as described in Fig. 4. Thus, $\Gamma \sim 500$ should be considered a lower limit. A more physical value for $Y_0$ must be greater than 1, making $\Gamma$ correspondingly larger and thus increasing the total isotropic energy of the burst at all energies. As shown in the Fermi LAT companion paper (*10*), the best estimate for the total apparent gamma-ray energy is $\sim10^{54}$ erg, which is uncomfortably large if the total bolometric energy inferred from $\Gamma$ is more than a factor of $Y_0 \sim 10$ larger.

| **Time Range** (s) | **Band $E_{peak}$** (keV) | **Band $\alpha_B$** | **Band C-Stat** (371 dof) | **Synchrotron $E_{peak}$** (keV) | **Electron Index $\delta$** | **Photospheric Energy $kT$** (keV) | **Total Flux 10 keV to 100 MeV** ($\gamma$ cm$^{-2}$ s$^{-1}$) | **Delta C-Stat Synch vs. Synch + BB** | **Synchrotron C-Stat** (/ dof) |
|---|---|---|---|---|---|---|---|---|---|
| **–0.1 : 0.0** | 5400 ± 1500 | –0.9 ± 0.1 | 330 | 2560 ± 1210 | 3.8 ± 0.5 | (no constraint) | 9 ± 1 | –0.8 | 348 / 383 |
| **0.0 : 0.1** | 1470 ± 170 | –0.49 ± 0.07 | 389 | 1800 ± 590 | 4.6 ± 0.6 | 276 ± 34 | 31 ± 3 | 39.1 | 433 / 381 |
| **0.1 : 0.2** | 849 ± 58 | –0.25 ± 0.06 | 365 | 1420 ± 320 | 4.8 ± 0.5 | 175 ± 10 | 46 ± 4 | 164. | 410 / 381 |
| **0.2 : 0.3** | 687 ± 38 | –0.33 ± 0.05 | 358 | 1270 ± 220 | 6.7 ± 1.5 | 146 ± 9 | 85 ± 5 | 130. | 398 / 381 |
| **0.3 : 0.4** | 553 ± 29 | –0.46 ± 0.04 | 369 | 975 ± 98 | 10.00 | 129 ± 9 | 116 ± 6 | 90. | 374 / 382 |
| **0.4 : 0.5** | 504 ± 26 | –0.57 ± 0.04 | 396 | 856 ± 60 | 10.00 | 86 ± 9 | 154 ± 6 | 43. | 395 / 382 |
| **0.5 : 0.6** | 380 ± 19 | –0.57 ± 0.04 | 364 | 640 ± 44 | 10.00 | 70 ± 6 | 152 ± 6 | 44. | 361 / 382 |
| **0.6 : 0.7** | 359 ± 20 | –0.64 ± 0.05 | 362 | 587 ± 39 | 10.00 | 54 ± 9 | 149 ± 6 | 12. | 376 / 382 |
| **0.7 : 0.8** | 327 ± 19 | –0.73 ± 0.05 | 362 | 560 ± 41 | 10.00 | 34 ± 5 | 137 ± 7 | 19. | 411 / 382 |
| **0.8 : 0.9** | 280 ± 18 | –0.76 ± 0.05 | 396 | 386 ± 30 | 10.00 | 53 ± 9 | 126 ± 6 | 22. | 390 / 382 |
| **0.9 : 1.0** | 253 ± 17 | –0.72 ± 0.06 | 370 | 338 ± 30 | 10.00 | 49 ± 11 | 106 ± 6 | 7.0 | 365 / 382 |
| **1.0 : 1.1** | 208 ± 16 | –0.76 ± 0.07 | 357 | 304 ± 25 | 10.00 | 30 ± 15 | 95 ± 6 | 2.0 | 380 / 382 |
| **1.1 : 1.2** | 207 ± 16 | –0.69 ± 0.08 | 327 | 347 ± 40 | 10.00 | 22 ± 3 | 74 ± 7 | 1.2 | 356 / 382 |
| **1.2 : 1.3** | 182 ± 13 | –0.73 ± 0.08 | 315 | 238 ± 27 | 10.00 | 43 ± 8 | 84 ± 5 | 20. | 350 / 382 |
| **1.3 : 1.4** | 170 ± 11 | –0.62 ± 0.09 | 393 | 273 ± 26 | 10.00 | 29 ± 4 | 85 ± 6 | 5.9 | 396 / 382 |
| **1.4 : 1.5** | 167 ± 12 | –0.69 ± 0.09 | 363 | 254 ± 26 | 10.00 | 24 ± 4 | 79 ± 7 | 12. | 380 / 382 |
| **1.5 : 1.6** | 165 ± 13 | –0.81 ± 0.09 | 342 | 233 ± 24 | 10.00 | 28 ± 5 | 73 ± 6 | 15. | 354 / 382 |
| **1.6 : 1.7** | 130 ± 9 | –0.5 ± 0.1 | 363 | 169 ± 20 | 10.00 | 31 ± 6 | 73 ± 5 | 13. | 356 / 382 |
| **1.7 : 1.8** | 142 ± 11 | –0.7 ± 0.1 | 349 | 193 ± 20 | 10.00 | 29 ± 5 | 77 ± 6 | 7.2 | 380 / 382 |
| **1.8 : 1.9** | 128 ± 10 | –0.5 ± 0.1 | 297 | 184 ± 20 | 10.00 | 27 ± 5 | 69 ± 6 | 7.6 | 321 / 382 |
| **1.9 : 2.0** | 137 ± 11 | –0.7 ± 0.1 | 339 | 199 ± 22 | 10.00 | 23 ± 4 | 68 ± 7 | 6.7 | 346 / 382 |
| **2.0 : 2.1** | 120 ± 10 | –0.7 ± 0.1 | 341 | 198 ± 23 | 10.00 | 21 ± 4 | 60 ± 7 | –0.8 | 317 / 382 |
| **2.1 : 2.2** | 114 ± 10 | –0.6 ± 0.1 | 366 | 136 ± 21 | 10.00 | 27 ± 4 | 51 ± 5 | 18. | 364 / 382 |
| **2.2 : 2.3** | 109 ± 11 | –0.7 ± 0.15 | 305 | 221 ± 41 | 10.00 | 17 ± 2 | 41 ± 6 | 5.4 | 300 / 382 |
| **2.3 : 2.4** | 109 ± 11 | –0.8 ± 0.2 | 299 | 112 ± 32 | 10.00 | 35 ± 8 | 46 ± 5 | 8.1 | 321 / 382 |
| **2.4 : 2.5** | 101 ± 11 | –0.7 ± 0.2 | 330 | 170 ± 39 | 10.00 | 13 ± 2 | 35 ± 7 | –2.4 | 358 / 382 |

**Table S1.** Spectral fit results for the onset pulse of GRB 130427A. Columns 2–4 contain the varying Band fit parameters ($\beta$ was fixed) and C-Stat fit statistic, while the rest of the columns are derived from the synchrotron plus blackbody function spectral fit. The synchrotron peak energy (column 5) has not been scaled by the square of the electron minimum Lorentz factor. After 0.3 s, the electron distribution power law index (column 6) could not be constrained by the data and was thus fixed to the steep value of 10 (essentially equivalent to thermal electrons). Column 9 indicates the change in the fitting merit function between the synchrotron model alone versus the synchrotron model plus blackbody. Although simulations are required to determine probabilities accurately, a value greater than 20 indicates that the extra blackbody component is required at the 4 sigma level, assuming normal statistics. Parameter uncertainties are 1 sigma statistical only. Times are relative to trigger time $T_0$.